\documentclass{aa501} 
 
\usepackage{graphicx,amssymb} 
\usepackage{natbib}

\begin{document} 
 
 
\title{Far-infrared dust opacity and visible extinction in the Polaris Flare} 
 
\titlerunning{Far--infrared dust opacity and visible extinction} 
 
\author{ 
L. Cambr\'esy\inst{1,2} \and 
F. Boulanger\inst{3} \and 
G. Lagache\inst{3} \and 
B. Stepnik\inst{3} 
}  
 
\authorrunning{L. Cambr\'esy et al.} 
 
\institute{ 
Infrared Processing and Analysis Center, Jet Propulsion Laboratory,
California Institute of Technology,
Mail Code 100-22, Pasadena, CA 91125 \and 
Observatoire de Paris, D\'epartement de Recherche Spatiale,  
F-92195 Meudon Cedex, France \and 
Institut d'Astrophysique Spatiale, Universit\'e Paris Sud, Bat. 121, 
91405 Orsay Cedex, France 
}  

\offprints{Laurent Cambr\'esy,\\ laurent@ipac.caltech.edu} 
             
\date{Received / Accepted} 
 
\abstract{ 
We present an extinction map of the Polaris molecular cirrus cloud derived
from star counts and compare it with the \citet{SFD98} extinction map
derived from the far--infrared dust opacity.
We find that, within the Polaris cloud, the \citet{SFD98}
$A_V$ values are a factor 2 to 3 higher than the star count values. 
We propose that this discrepancy results from a difference in  
$\tau_{\rm FIR}/ A_V$ between the diffuse atomic medium and the Polaris 
cloud. We use the difference in spectral energy distribution, {\em warm} for 
the diffuse atomic medium, {\em cold} for the Polaris cloud, to separate
their respective contribution to the line of sight integrated infrared
emission and find that the $\tau_{\rm FIR}/ A_V$ of {\em cold} dust in
Polaris is on average 4 times higher than the \citet{SFD98} value for dust
in atomic cirrus. This change in dust property could be interpreted by a
growth of fluffy particles within low opacity molecular cirrus clouds such
as Polaris.  
Our work suggests that variations in dust emissivity must be taken  
into account to estimate $A_V$ from dust emission wherever {\em cold}
infrared emission is present (i.e. molecular clouds). 
}

\maketitle

\section{Introduction} 
The infrared sky images provided by the Infrared Astronomy Satellite 
(IRAS) and the Cosmic Background Explorer (COBE) have 
considerably improved our knowledge of interstellar dust and of the 
spatial distribution of interstellar gas at high Galactic latitude. 
With the IRAS data it became clear that interstellar dust comprises 
small particles stochastically heated by the absorption of photons 
to temperatures higher than the equilibrium temperature of large grains 
\citep{DBP90}. 
Far from heating sources, these small particles make the 12, 25 and a 
significant fraction of the 60~$\mu$m IRAS emission.  
Only the 100~$\mu$m emission comes from large grains emitting  
at the equilibrium temperature sets by the balance 
between heating and cooling. COBE extended the 
IRAS observations to the far--infrared/sub--millimeter emission. 
These data have been used to measure large grain
properties (temperature and emissivity) and trace the distribution 
of interstellar matter \citep{WMB+91,RDF+95,BAB+96,LABP98,AOW+98,SFD98,FDS99}.  
In particular, \citet{SFD98} (hereafter SFD98)  
combined  the angular resolution of IRAS 100~$\mu$m data  
and the wavelength coverage of DIRBE to make  
a map of the  dust far--infrared (FIR) opacity. 
They have shown that, in regions of low extinction ($A_V \leq 0.3$) at 
high Galactic latitude, the FIR dust optical depth, 
$\tau_{\rm FIR}$, is tightly correlated with visible extinction. 
By extrapolating this correlation to the whole sky they produced an all-sky
map of visible extinction which is being used for a wide range of purposes. 
 
The IRAS and COBE observations show that the spectral energy distribution of
dust emission varies in the ISM and in particular from the diffuse 
atomic medium to molecular clouds \citep[e.g.][]{LCP91,ABMF94, LABP98}.
These variations have been interpreted 
in terms of changes in dust composition, in particular variations in 
the abundance of small grains. What is still a matter of debate is 
whether they also imply changes in the properties of large grains. 
In this paper we address this question by looking for variations 
in the $\tau_{\rm FIR}/ A_V $ ratio.   
In a first part of the paper, we compare the SFD98 extinction map with 
an independent estimate of $\rm A_V$ derived from star counts over the 
Polaris Flare, a high latitude cirrus with CO emission and an extinction 
around 1 mag \citep{HT90,BAR+99}. 
We find that the SFD98 extinction is on average twice that estimated 
from star counts. In the second part, we investigate various explanations 
of the extinction discrepancy and propose to relate it to a difference 
in $\tau_{\rm FIR}/A_V$ between dust in low extinction cirrus and the 
colder dust associated with the Polaris Flare.

\section{Extinction in the Polaris Flare} 
 
\subsection{Extinction map from star counts} 
 
\begin{figure}[htb] 
	\includegraphics[width=8.8cm]{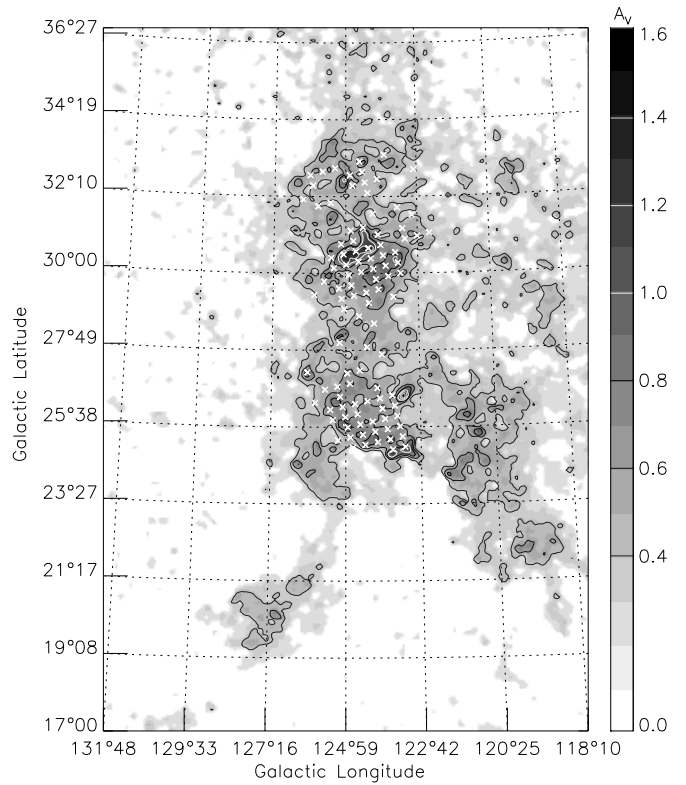} 
\caption{Visual extinction map of the Polaris Flare from $B$ star counts 
using the USNO-PMM catalogue. White crosses represent the DIRBE pixels used 
to compare the star count based extinction map with the SFD98 extinction map 
based on FIR emission} 
\label{polarisB} 
\end{figure}

We have used the USNO-PMM 
\citep{Mon96} $B$ photometry and the method described in \citet{Cam99a} to
build a visual extinction map (assuming $R_V = 3.1$) of the Polaris cirrus
cloud (Fig.~\ref{polarisB}). The method consists of star counts with variable
resolution  in which the local stellar density is estimated with the 20
nearest stars of each position. The extinction is derived from the star
density (D) as follows: 
\begin{eqnarray}
A_B &=& \frac{1}{a}\log \frac{D_{\rm ref}(b)}{D} \label{eq1}\\
A_V &=& A_B \times \frac{A_V}{A_B}= A_B \times \frac{R_V}{1+R_V}
\end{eqnarray}
where $a$ is the slope of the $B$ luminosity function and $D_{\rm ref}$
the reference stellar density in the absence of extinction.   
The high galactic latitude of the Polaris Flare ($b \sim 27\degr$) 
prevents the contamination by other clouds on the same line of sight. 
The upper limit to its distance of about 240 pc \citep{HSd+93} ensures that the 
cloud is close enough to derive the extinction from star counts without 
significant contamination by foreground stars.

The extinction map (Fig.~\ref{polarisB}) is obtained by taking for
$\log D_{\rm ref}(b)$ the value derived from the linear fit of $\log D$
versus the galactic latitude $b$ outside the Polaris Flare, i.e.
$\log D_{\rm ref} = \beta + \alpha \times b$.
This corrects the stellar density variations due to variations in the length
of the line of sight through the stellar disk, but ignores extinction from
diffuse interstellar matter outside the Polaris cloud.
To account for this diffuse uniform or slowly varying extinction which
is not {\em seen} by star counts, we measured the SFD98 map variations with
respect to the galactic latitude for regions with ($A_V < 0.2 $). 
For $b$ ranging from $25^\circ$ to $34^\circ$ 
and $l$ ranging from $105^\circ$ to $143^\circ$ we find that  
$A_V({\rm SFD98})-A_V({\rm starcount})\approx0.20-7.3\,10^{-3}\times b$. 
We have used this relation to correct the star count extinction
values in order to make the comparison with the SFD98 map; the absolute value
of the resulting correction is lower than 0.05 mag. After this correction,
both our extinction map and the SFD98 map measure the total extinction. This
is necessary to compare each map together and with DIRBE far--infrared data.
Extinction values outside the Polaris cloud (where $A_V \leq 0.2$) match those
of the SFD98 map.
For $17<b<25$ the linear relation we measured is no longer valid and it
becomes hard to make such a correction because of the lack of areas with low
extinction in the SFD98 map.
Since we are interested only in directions represented by white crosses
which are all at $b>25$ in Fig.~\ref{polarisB}, there is no need to try
to include diffuse extinction for $b<25$.
The star count Polaris extinction map agrees with extinction values
derived from color excesses based on CCD images of a 1 deg$^2$ section of
Polaris \citep{ZBB99}. 
 
The resolution of the resulting map is about 8\arcmin. Statistical 
uncertainties in the determination of the extinction come from star counts.
The distribution of stars follows a Poisson law with a sigma equal to
$\sqrt{N}$, where $N$ is the number of counted stars. For $N=20$ we obtain
$\Delta A_V = \pm^{0.29}_{0.23}$.  
Moreover, the $R_V$ parameter introduces a systematic uncertainty. In the
absence of any measurement, we have assumed $R_V=3.1$ which corresponds to the
average value in the diffuse interstellar medium. A higher value of 5.5,
that is common but not systematic in dense cores, would multiply our
extinction values by a factor 1.12.

\subsection{Comparison with SFD98 } 
\label{add_evid} 
\begin{figure}[htb] 
	\includegraphics[width=8.8cm]{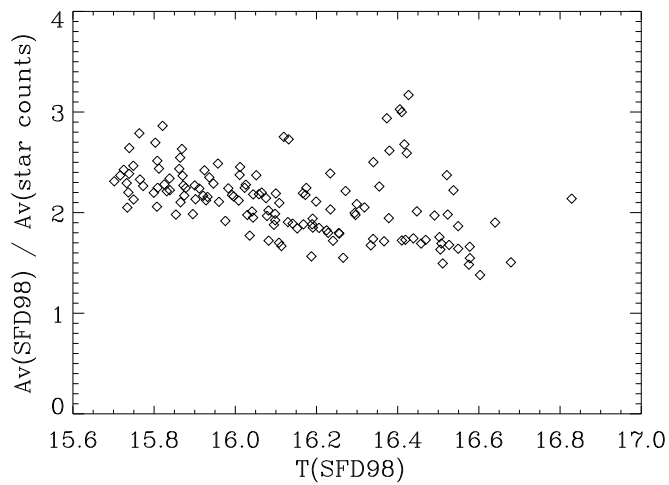} 
\caption{Comparison of the SFD98 extinction with star count extinction in 
the Polaris Flare versus the SFD98 temperature} 
\label{count_schl} 
\end{figure}

\begin{figure}[htb] 
	\includegraphics[width=8.8cm]{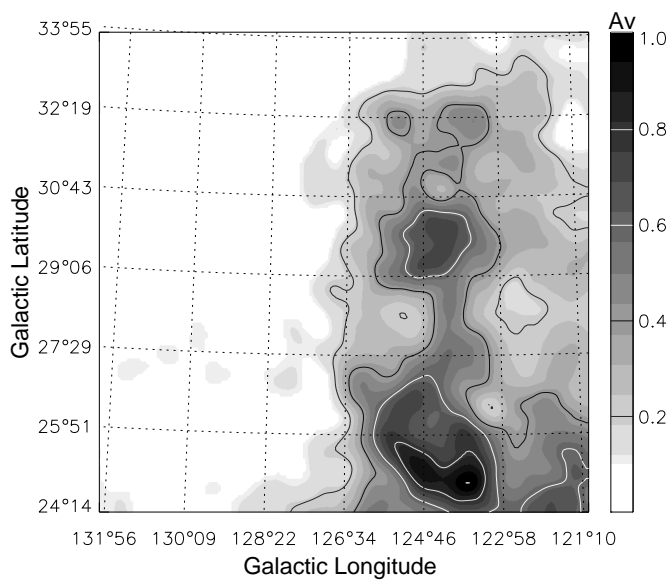} 
\caption{Difference between SFD98 and the star count extinction map
at the DIRBE resolution. The two maps are in agreement outside the molecular
cloud}
\label{zm_sc-av} 
\end{figure}

In Fig.~\ref{count_schl}, the ratio between SFD98 and star count
extinction values is plotted against the dust temperature as derived 
by SFD98. For this we have smoothed both extinction maps to the DIRBE  
resolution. The DIRBE pixels used for this comparison are marked with 
white crosses in Fig.~\ref{polarisB}. 
The SFD98 $A_V$ values are systematically higher than those derived from 
star counts. The extinction ratio decreases for increasing dust temperatures. 
The mean ratio between the SFD98 and star count extinction values is 2.1.
Fig.~\ref{zm_sc-av} shows the spatial distribution of the difference between
SFD98 and our extinction map.
 
The Polaris star counts raise a question about the validity of the SFD98
extinction map in regions of higher extinction that in those where they
calibrated their conversion from FIR dust opacity to $A_V$. 
Previous studies have already pointed out similar discrepancies. 
\citet{SG99} have presented a study of cirrus clouds 
with $UBVRI$ photometries and have compared their extinction derived from star 
counts, color excess, $UBV$ color--color diagrams and $BVI$ color--color 
diagrams. They found for their highest extinction cirrus, in Corona 
Australis, a discrepancy with the SFD98 map comparable to  
that found here for the Polaris Flare.  
\citet{AG99a} have compared the extinction toward the Taurus 
cloud using four different techniques and they also concluded 
that the  SFD98 map overestimates the extinction by a factor of 1.3-1.5.  
These results are confirmed by \citet{vM01} who compared a color study
of the globular cluster NGC~3201 ($b=8.6^\circ$) with the SFD98 extinction
map for this line of sight. They concluded their analysis on a reddening
overestimation in the SFD98 map.

\subsection{Discussion} 

The extinction data presented in this paper together with similar data on 
other molecular clouds all lead to the same conclusion: the SFD98 extinctions 
are larger than the values derived from star counts in molecular clouds where
$A_V \ge 1$ .  

Cloud structure on angular scales smaller than the resolution of the 
star counts can make the measured $A_V$ smaller than the mean value of $A_V$. 
The relation between the mean extinction and the stellar density is not
linear but logarithmic (Eq.~\ref{eq1}) whereas the FIR integration over a
cell is linear.
This could thus explain the extinction discrepancy.
\citet{Ros80} has quantified this effect. Using his results, we found   
that a difference of a factor 2 for a visual extinction of 2 magnitudes 
would require a surface filling factor lower than 0.1.
Such a low surface filling factor is incompatible with studies of the
clumpiness of dust extinction which conclude on a {\em smooth} distribution
of the dust in molecular clouds \citep{TBD97,LAL99}.
 
\begin{figure}[htb]
        \includegraphics[width=8.8cm]{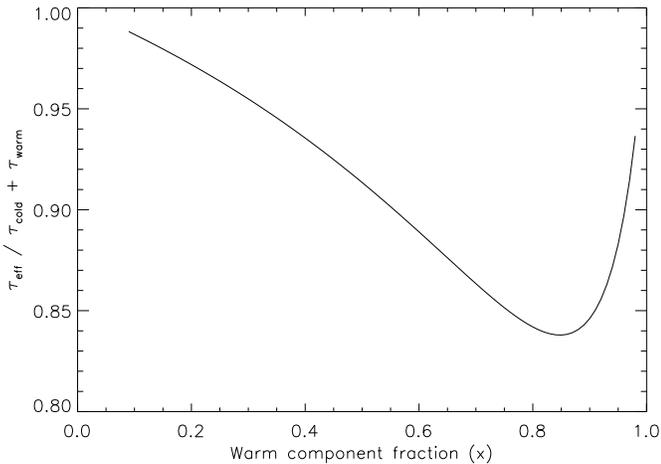}
\caption{Effective to total optical depth ratio versus {\em warm} component
        fraction. {\em Cold} and {\em warm} components are described by the
        modified Planck function
        $P = B_{\nu}(T) \times (\lambda/ \lambda_0)^{-2}$
        with $T_{\rm c}=13.5$~K and $T_{\rm w}=17.5$~K, respectively.
        Optical depths are derived by fitting each spectrum with a
        single modified Planck function.
}
\label{tau_eff}
\end{figure}

Another plausible explanation would be a shortcoming in the
SFD98 dust temperature determination in molecular clouds 
which ignores temperature variation. 
This should be considered due to the coarse resolution of the
SFD98 temperature map ($\sim 1^\circ$) compared to the clumpiness of the
brightest structures in cirrus clouds. For example,
for a given FIR brightness, an underestimation of the dust temperature 
translates into an overestimation of the dust opacity. 
Observations of molecular clouds at higher angular resolution than DIRBE  
do show temperatures locally lower than those of SFD98. For example, PRONAOS 
observations of a piece of the Polaris cirrus gave a dust temperature of
13.5~K \citep{BAR+99}, significantly lower than the value of 16~K at the same
position in the SFD98 temperature map. This example clearly shows that the
SFD98 temperature is an effective temperature which represents a mean intensity 
value weighted over the dust seen within the solid angle of their study. 
The impact of the beam (and line-of-sight) averaging on the SFD98 temperatures and opacities 
can be discussed on the basis of a simple model where we assume that the
emission comes from a mixture of two distinct components. In this calculation,
we assume an emissivity law ($\propto \nu^2$) and fit a single black body to a
combination of two black bodies at different temperatures. The effective optical
depth is always smaller than the sum of the optical depth for the two
temperatures. Thus if the $\tau_{\rm FIR}/A_V$ is the same for both dust
components, our simple model shows that the SFD98 method always lead to an
underestimation of $A_V$, the reverse of the discrepancy observed with star
counts (Fig.~\ref{tau_eff}). 
More generally, when a range of temperatures is present within the beam, 
the effective FIR dust opacity that is measured by SFD98 is always
smaller than the mean opacity. Temperature variations within the beam thus
cannot explain the extinction discrepancy between SFD98 and star counts.  

The SFD98 map is a $\tau_{\rm FIR}$ map and not an $A_V$ map. To go from one
to the other one needs to know the $\tau_{\rm FIR}/A_V$ ratio. We propose to
explain the extinction discrepancy by variations in the $\tau_{\rm FIR}/A_V$
ratio between the low extinction regions used for the calibration of the $A_V$
and the Polaris molecular cirrus. 
Variations in the $R_V$ ratio
from diffuse to dense clouds are interpreted as evidence for variations in the
optical properties of large dust grains in molecular clouds \citep{CCM89}.
In our interpretation, the extinction
discrepancy would thus be an additional signature of
the evolution of dust from diffuse to dense clouds.

\section{{\em Warm} and {\em cold} dust components} 
 
\subsection{Evidence for distinct emission components} 
 
The IRAS data have shown that no emission  from aromatic hydrocarbons  
and very small grains at 12, 25 and 60~$\mu$m is seen towards the dense gas
traced by the millimeter transitions of $^{13}$CO \citep{LCP91,ABMF94}.
In practice, the molecular emission from this dense gas is observed to be
well correlated with the difference between the 100 and 60~$\mu$m IRAS
brightnesses.   
\citet{LABP98} combined the FIR DIRBE bands to the 60~$\mu$m 
data to separate along each line of sight the emission associated with the
diffuse ISM and dense gas where both are present. With the long wavelength
bands of DIRBE (100, 140 and 240~$\mu$m), they were able to determine a dust
temperature for each of these two emission components. Their results show that
the dust associated with the dense gas, as traced by the difference between
the 100 and 60~$\mu$m brightness, is systematically colder ($T \sim 15$~K)
than dust in the diffuse atomic interstellar medium ($T \sim 17.5$~K). We
refer to these two emission components as {\em cold} and {\em warm}.  
The distinction between {\em cold} and {\em warm} emission components
characterized by different dust temperatures and the abundance
of small grains emission is corroborated by sub-mm observations at higher
angular resolution for which the spatial separation of the infrared emission
from diffuse and dense gas is easier to distinguish than with the DIRBE data
\citep{BAR+99,SAB+01, LHB+96}.  

We propose to relate the extinction discrepancy to a variation in the
$\tau_{\rm FIR}/A_V$ ratio  between the {\em cold} and {\em warm} dust
components. 
In their analysis of the DIRBE data, SFD98 did not consider the 
possible presence of distinct emission components with different dust
temperature along the line of sight. They assume that large grain properties
are everywhere the same and consequently that the temperature variations are
exclusively due to changes in dust heating. 
In the \citet{LABP98} description of the data, the observed variations in the
dust effective emission temperature are to a large extent due to various
degrees of mixing between the {\em warm} and {\em cold} components.
In a second paper, \citet{FDS99}, the same authors as SFD98, showed the
necessity of including a very cold emission component with $\rm T \sim 9 K$
in their analysis to improve the fitting of the FIRAS data at long 
wavelengths. However, this very cold component and the {\em cold} component of
\citet{LABP98} are distinct (Note that they are not mutually exclusive
and that both may be required to correctly describe the data). 
The {\em cold} component, derived from DIRBE data, is spatially correlated with known
molecular clouds; the very cold component, deduced from FIRAS long-wavelength spectra is, 
on the contrary, present in all directions.
For the subject of this paper the key difference between both components is that in the
\citet{FDS99} analysis the very cold component accounts for a fixed fraction
of the emission, independent of sky position (fixed opacity ratios between the
very cold and warm components, with temperature given by a power law) while in
the \citet{LABP98} analysis the {\em cold} fraction varies spatially and is zero
in the high latitude regions used by SFD98 for the calibration of the
$\tau_{\rm FIR}/A_V$ ratio. Due to this spatial dependence, a difference in
dust properties between the {\em cold} and {\em warm} emission components,
namely in the $\tau_{\rm FIR}/A_V$ ratio, will limit the domain of the SFD98
conversion of $\tau_{\rm FIR}$ into $A_V$ and could account for the observed
discrepancy with star counts. This statement is quantified in the following
sections.

\subsection{Separation of the {\em warm} and {\em cold} components} 
\label{separation} 
 
A separation of the {\em warm } and {\em cold} contributions to the
infrared brightness is necessary in order to derive a temperature and FIR
opacity for each component. To do this, we apply a method similar to that of
\citet{LABP98} to DIRBE images of the Polaris Flare. 
Let $I(\lambda)$ ($\lambda$=100, 140, 240~$\mu$m) be the DIRBE emission 
at 100, 140 and 240~$\mu$m respectively, and $R(\lambda,60)$ the 
$I_{\nu}(\lambda)/I_{\nu}(60)$ flux ratio observed in cirrus clouds. The 
{\em cold} maps are computed at each wavelength according to the relationship: 
$I(\lambda)_{\rm c}=I(\lambda)-R(\lambda,60)\times I(60)$.\\ 
 
We first determine $R(100,60)$ using the correlation diagram of the diffuse
cirrus emission around the Polaris cloud.  We obtain $R(100,60) = 4.0 \pm 0.3$.
At 140 and 240~$\mu$m the correlation diagrams are much more noisy. Therefore
we prefer to compute $R(140,60)$ and $R(240,60)$ using $R(100,60)$ and assuming
a large grain temperature of 17.5$\pm$0.5~K. We thus obtain:
$R(140,60) = 7.9 \pm 1.1$ and $R(240,60)=6.4 \pm 1.4$.\\ 
 
The produced {\em cold} maps have non-zero emission at a large scale due to 
1) zodiacal residual emission (mainly coming from the 60~$\mu$m map) and 2) 
non-correction of the Cosmic Infrared Background. 
Therefore, under the assumption that the {\em cold} dust is distributed  
with limited angular extent in molecular clouds, we remove the low frequency 
structures using a $18\degr \times 18\degr$ median filter. We thus obtain, 
at the end, maps of {\em cold} dust emission at 100, 140, and 240~$\mu$m 
which are clumped on scales smaller than our filter size. 
 
Map of statistical uncertainties for the {\em cold} emission are computed using 
the DIRBE release error maps: 
\begin{equation} 
\Delta_\mathrm{sta} I_{\nu}(\lambda)_{\rm c} =  
\sqrt{\Delta I_{\nu}(\lambda)^2 + R(\lambda,60)^2 \times \Delta I_{\nu}(60)^2} 
\end{equation} 
Systematic uncertainties are derived following: 
\begin{equation} 
\Delta_\mathrm{sys} I_{\nu}(\lambda)_{\rm c} = 
\Delta R(\lambda,60)\times I_{\nu}(60) 
\end{equation} 
Temperatures and optical depths are derived only for pixels containing 
significant {\em cold} emission (see Fig.~\ref{polarisB}), i.e. {\em cold} 
intensities at 100, 140 and 240~$\mu$m greater than 3$\sigma$, $\sigma$ 
being estimated at each wavelength using the width of the histograms of the 
{\em cold} emission maps. 
 
Statistical and systematic uncertainties on the {\em cold} dust temperature and 
optical depth are computed using the statistical and systematic 100, 140 and 
240~$\mu$m error maps of {\em cold} emission, respectively. Temperatures and 
optical depths are derived using the $\chi^2$ fitting and assuming a 
FIR dust emissivity index of 2. Uncertainties given by the $\chi^2$
fitting correspond to the 68.3\% confidence level.\\ 
 
To determine the {\em warm} optical depth, we adopt the following method. 
We first remove to the 100~$\mu$m map the {\em cold} 100~$\mu$m emission and 
the cosmic FIR background value from \citet{LHRT00}. 
We then assume that the {\em warm} component has a temperature of 
17.5$\pm$0.5~K with a FIR dust emissivity index of 2. Optical
depth is thus directly proportional to the 100~$\mu$m {\em warm} emission.  
 
The error on the {\em warm} emission, including both the systematic 
and statistical errors, is: 
\begin{equation} 
\label{err_warm} 
\Delta I_{\nu}(100)_{\rm w} = \sqrt{\Delta I_{\nu}(100)^2 + 
\Delta I_{\nu}(100)_{\rm c}^2 + \Delta I_{\nu}(100)_{\rm CIB}^2} 
\end{equation} 
Optical depth uncertainties correspond to the minimum and maximum optical depth 
value allowed by the combination of  $\Delta I_{\nu}(100)_{\rm w}$ and the 
0.5~K error on the assumed dust temperature.

\subsection{From far--infrared optical depth to visible extinction} 
\label{method} 
 
Within the two components model, the total extinction can be written as the
sum of the extinction in the {\em warm} and in the {\em cold} components: 
\begin{equation} 
A_V = A_V^{\rm w} + A_V^{\rm c}  
\end{equation} 
and we have: 
\begin{eqnarray} 
A_V^{\rm w} = \tau_{100}^{\rm w} \times \left(\frac{\tau_{100}}{A_V}\right)_{\!\!\rm w}^{-1}\label{av_w}\\ 
A_V^{\rm c} = \tau_{100}^{\rm c} \times \left(\frac{\tau_{100}}{A_V}\right)_{\!\!{\rm c}}^{-1} 
\end{eqnarray} 
where the ratio $(\tau_{100} / A_V)_{\rm w,c}$ are the emissivities of the dust 
at 100~$\mu$m for the {\em warm} and {\em cold} components, respectively. 
We take for the FIR-to-visible opacity ratio of the {\em warm}
component, $(\tau_{100} / A_V)_{\rm w}$, the ratio measured by SFD98 in low
extinction regions. Then, we can derive $A_V^{\rm w}$ using Eq.~\ref{av_w}
and the emissivity of the {\em cold} component is: 
\begin{equation} 
\left(\frac{\tau_{100}}{A_V}\right)_{\!\!\rm c} = \frac{\tau_{100}^{\rm c}}{A_V - A_V^{\rm w}} 
\label{rescold} 
\end{equation}

\subsection{Far--infrared emissivity of the {\em warm} component} 
 
The dust emissivity for the {\em warm} component is taken from SFD98. In their 
paper, they provide the value of $p=E_{B-V}/I_{100}=0.0184$, where $I_{100}$ 
is the 100~$\mu$m brightness expressed in MJy~sr$^{-1}$ corrected from the
zodiacal emission and scaled with the measured dust opacity from the pixel
dependent dust temperature to a fixed temperature of 18.2~K. With this
scaling, the conversion from $I_{100}$ to FIR opacity becomes pixel independent.
We have: 
\begin{equation} 
\frac{\tau_{100}}{A_V}= \frac{I_{100}}{A_V \times \mathcal{B}_{100}(T)} = \frac{1}{R_V \, p \, \mathcal{B}_{100}(18.2\mathrm{K})} 
\end{equation} 
For the mean Solar Neighborhood extinction law we have $R_V=A_V/E_{B-V}=3.1$. 
Moreover, a color correction should be applied to the optical depth because 
FIR fluxes are expressed for the effective filter wavelength
assuming a $\nu^{-1}$ law. The correction consists of dividing the infrared
flux by a color factor $\mathcal{K}$ defined in the DIRBE {\em explanatory
supplement} as follow: 
\begin{equation} 
{\cal K}=\frac{{\displaystyle \int_\nu} 
        \left(\frac{I_\nu}{I_{\nu_0}}\right)_{\!\!\rm real} W_\nu\, d\nu} 
       {{\displaystyle \int_\nu} 
        \left(\frac{I_\nu}{I_{\nu_0}}\right)_{\!\!{\rm eff}} W_\nu\, d\nu} 
\end{equation} 
where $W_\nu$ is the filter profile. The value given in the {\em explanatory
supplement} is $\mathcal{K}_{100}(18.2K) = 0.91125$. 
 
Finally, the emissivity for the {\em warm} component is: 
\begin{equation} 
\left(\frac{\tau_{100}}{A_V}\right)_{\!\!\rm w} = \frac {{\cal K}_{100}(18.2\textrm{K})}{R_V\, p\, {\cal B}_{100}(18.2\textrm{K})} = 1.31\,10^{-7} 
\end{equation}

\subsection{Far--infrared emissivity of the {\em cold} component} 
\subsubsection{Result} 
Following the procedure described in Sect.~\ref{separation} and \ref{method} 
we use the extinction map (Fig.~\ref{polarisB}) and the decomposition of the
FIR flux in two components. We use the DIRBE resolution (the
extinction map has been convolved by the DIRBE beam) and only pixels with
high signal-to-noise ratio for both components. 
\begin{figure*}[htb] 
	\includegraphics[width=18cm]{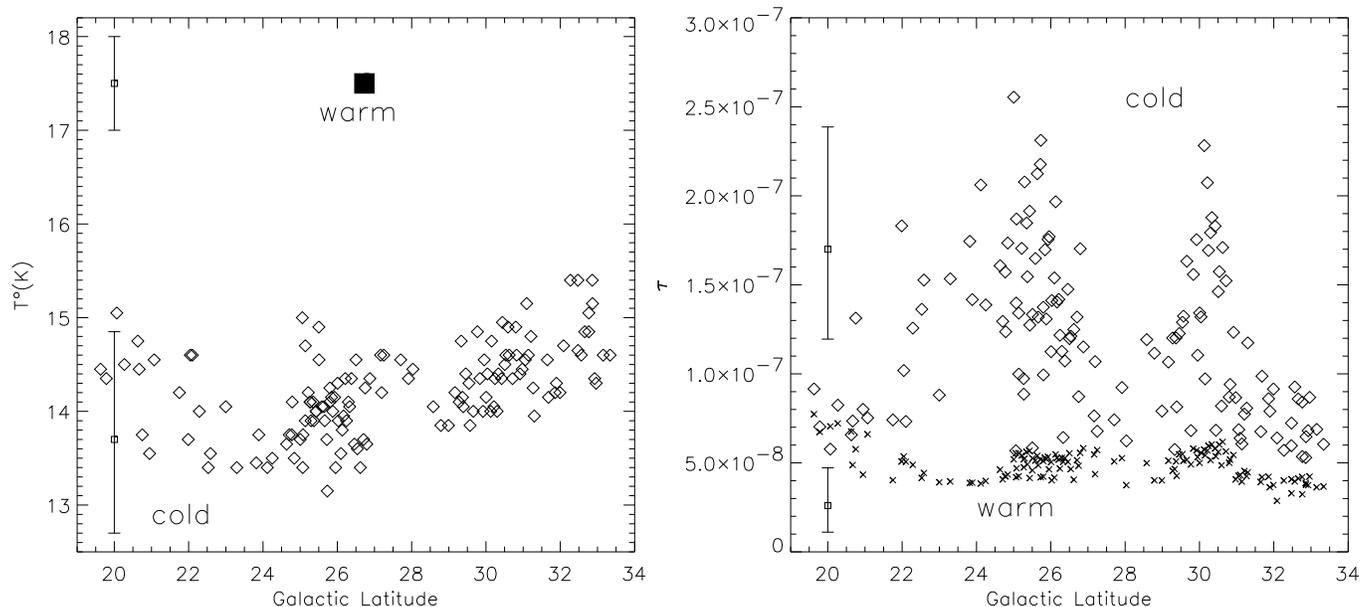} 
\caption{Temperature and optical depth normalized at $1 \mathrm{cm}^{-1}$ 
versus the galactic latitude for {\em warm} and {\em cold} components. Error 
bars represent the mean total uncertainty (statistical plus systematic) for 
each component} 
\label{warmcold} 
\end{figure*} 
Fig.~\ref{warmcold} show the temperature and the optical depth normalized to 
$1 \mathrm{cm}^{-1}$ for the two components. The temperature in the Polaris 
Flare varies from 13~K to 15.5~K. 
 
\begin{figure}[htb] 
	\includegraphics[width=8.8cm]{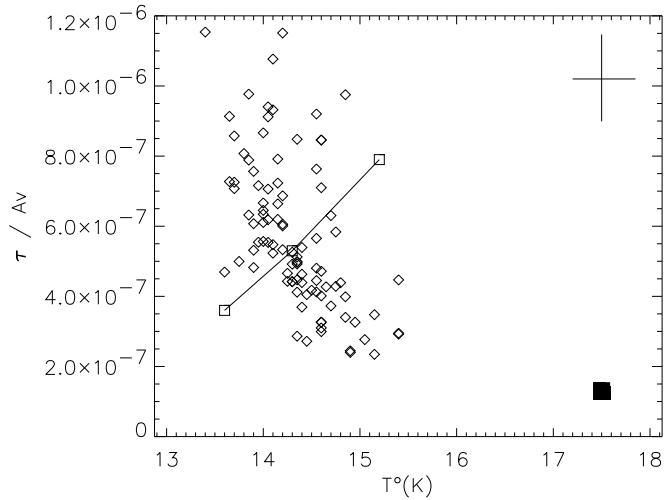} 
\caption{Emissivity variation versus temperature in the Polaris Flare, 
with the typical statistical (cross in the upper right corner) and systematic
(squares) uncertainties. The filled square represents the SFD98 value which
corresponds to the {\em warm} component emissivity. Dust emissivity of the
{\em cold} component is 4.0 times larger than the {\em warm} one} 
\label{result} 
\end{figure} 
Using Eq.~\ref{rescold} we obtain the values of the emissivity of the 
{\em cold} component, $\left(\frac{\tau_{100}}{A_V}\right)_{\!\!\rm c} $,  
presented in the Fig.~\ref{result}. These values are all above the SFD98 
value for the {\em warm} component (filled square in Fig.~\ref{result}).  
The median value for the {\em cold} component emissivity at 
100~$\mu$m is: 
\begin{equation} 
\left(\frac{\tau_{100}}{A_V}\right)_{\!\!\rm c} = 5.3 \, 10^{-7} 
\end{equation} 
 
This is 4.0 times larger than the SFD98 value for the {\em warm} component. 
 
\subsubsection{Uncertainties} 

In this sub-section, we compare  uncertainties on the dust parameters with 
the dispersion of the points in Fig.~\ref{result}. For the {\em warm}
component, we are not able to separate the systematic from the statistical 
errors (see Eq.~\ref{err_warm}). Since  these errors are small compared to
those on the {\em cold} component parameters (by a factor of $\sim 3$) we
focus on this last contribution to the observed scatter in  Fig.~\ref{result}.

\paragraph{Systematic errors.} 
In Sect. \ref{separation} we describe errors associated with the separation of 
the FIR flux in two components. Systematic errors on 
$\tau_{100} / A_V$ are dominated by uncertainties on $R(\lambda,60)$ which 
are used to derive the emission of the {\em cold} dust. 
These systematic errors would shift the whole set of data points.  
In Fig.~\ref{result} we give the direction and amplitude of these 
shifts for the median value. Extreme values for  
$\left(\frac{\tau_{100}}{A_V}\right)_{\!\!\rm c}$ are $3.6\,10^{-7}$ and 
$7.9\,10^{-7}$ and correspond to 2.7 and 6.0 respectively for the ratio of 
the emissivity between the two components. 
For $R_V=5.5$, we have already mentioned that star count extinction is
multiplied by 1.12; the resulting median emissivity ratio would then be 3.1
(rather than 4.0).
 
\paragraph{Statistical errors.} 
The dispersion of the values represented in Fig.~\ref{result} suggests 
an increase of the FIR emissivity of the dust when the temperature 
decreases. However we examine the possibility of this trend being linked with
statistical errors. In Fig.~\ref{result} the statistical errors are represented
by error bars in the upper right corner. 
Since the optical depth is related to the FIR emission by the 
relation $\tau_{100} = I_{100}/\mathcal{B}_{100}(T)$, errors on the temperature 
and on the optical depth are correlated: temperature overestimation implies 
optical depth underestimation. 
This effect contributes to the observed dispersion in Fig.~\ref{result} 
but cannot fully explain its amplitude. We believe therefore that an 
intrinsic dispersion of the $\tau_{100}/A_V$ exists and that it may be 
related to an evolution of the emissivity with the temperature.  
 
\section{Discussion} 
\label{discussion} 
 
\subsection{Dust evolution from {\em warm} to {\em cold} component} 
 
The submillimeter analysis of the Polaris Flare by \citet{BAR+99} has
revealed an unexpected low temperature in this cirrus. They have used the
extinction map of Fig.~\ref{polarisB} to model the effect of the
attenuation of the interstellar ultraviolet and visible radiation field
on the dust temperature.
They concluded that this cannot explain the observed temperature difference
between Polaris and {\em warm} cirrus clouds. They thus suggest that the
temperature difference is related to a change in optical dust properties. 
For a larger FIR to visible/UV extinction ratio, grains are able to radiate
more efficiently the energy absorbed and thus their equilibrium temperature
is reduced. Thereby, our interpretation of the extinction data is in agreement
with \citet{BAR+99} conclusions. A similar conclusion about the FIR dust
emissivity is reached by the modeling of PRONAOS FIR/sub-mm observations of
a molecular filament in Taurus \citep{SAB+01}.
In this modeling, a change in the $\tau_{\rm FIR}/A_V$ ratio between
the {\em cold} dust within the filament and the {\em warm} dust outside
by a factor 3.4 is required to reproduce the brightness profiles across the
filament at the various wavelengths. This study, unlike, ours does not
rely on the empirical separation of {\em warm} and {\em cold} components.

The most straightforward explanation for the change in dust FIR emissivity
between diffuse and molecular clouds is grain growth through grain-grain
coagulation and accretion of gas species \citep[e.g.][]{Dra85a}. These two
processes should lead to composite grains with significant porosity. 
The effect of the formation of composite fluffy grains on the dust opacity is
well explained by \citet{Dwe97}. Two effects contribute to an increase
in the dust opacity per unit dust mass. 
(1) The first contribution results from an increase in the effective grain
sizes. For spherical grains of size $a$:
$\tau_{\rm ext} = \pi a^2 Q_{\rm ext} N_d$ where $N_{\rm d}$ is the dust
column density and $Q_{\rm ext}$ the extinction efficiency, and 
$M_{\rm d} = 4/3 \pi a^3 \rho N_{\rm d}$, where $\rho$ is the grain density.
Within the Rayleigh limit ($2 \pi a \ll \lambda$), $Q_{\rm ext}/a$ is
independent of the grain size. For particles large relative to the wavelength
($2 \pi a \gg \lambda$), it is $Q_{\rm ext} $ which is roughly size-independent.
In practice, the Rayleigh limit applies to all grains in the FIR while the
second limiting case applies to large interstellar grains ($a\sim 0.1 \mu$m)
in the UV. Within these limiting cases, one finds that $\tau_{\rm FIR}$ scales
as $1/\rho$ while $\tau_{\rm UV}$ scales as $1/a \rho$. For increasing grain
porosity, the effective grain density $\rho$ decreases and $\tau_{\rm FIR}$
increases. Thus, the ratio $\tau_{\rm FIR} / \tau_{\rm UV}$ scales as the
grain size $a$. Qualitatively this also applies to the ratio
$\tau_{\rm FIR} / A_V $.
(2) The second, more subtle, effect is that the optical properties of the
composite grains differ from the optical properties of their constituents: 
grain properties depend not only on their size but also on their
composition and structure. In particular, the wavelength dependence of
$Q_{\rm ext}/a$ changes with grain composition. For the specific case of
composite carbon-silicate grains, this leads to a significant enhancement
of the opacity in the FIR relative to that in the visible \citep{Dwe97}. 
Dust properties of porous composite grains have also been quantified
numerically \citep[e.g.][]{BD90,MW89}. In their work, \citet{MW89}
show that it is possible to fit the extinction curve and its variations
with $R_V$ with a size distribution of composite porous carbon+silicate
grains. Their calculations show that enhancements of the $\tau_{\rm FIR} / A_V $
ratio by a factor of at least 3 can be obtained within the constraints set
by the UV to near--infrared extinction curve.
 
The proposed interpretation of the variations in FIR dust emissivity implies
that grain coagulation would be effective even within cirrus clouds and not
only in dark clouds and proto-stellar condensations. Observed variations in
the $R_V$ ratio have  also been interpreted as evidence for grain growth
from diffuse to dense clouds \citep[e.g.][]{KM96}. It will thus be interesting
to look for a plausible correlation between the changes in $\tau_{\rm FIR}/A_V$
and variations in the visible extinction curve. Note however that these two
signatures of dust evolution from the diffuse ISM to molecular clouds are not
necessarily correlated because $R_V$ and $\tau_{\rm FIR}/A_V$ do not share the
same dependence on the dust porosity and size distribution \citep{MW89}.

In the absence of multiband visible photometric data, we have no estimate of
$R_V$ in the Polaris cloud and one can only make rather qualitative comments. 
\citet{CCM89} have shown that the variations in the extinction curve from the
UV to the near-UV, to a good approximation, can be related to one single
parameter, $R_V$. The $R_V$-dependent extinction curves all converge in the
near-IR but the change in the $\tau_{\rm FIR}/A_V$ ratio proposed to explain
the extinction data is larger that one would expect from the simple
extrapolation of this convergence  to the FIR.
For the \citeauthor{CCM89} relation, the $\tau_{\rm FIR}/A_V$ varies by a
factor 1.2 for a large change in $R_V$ from 3.1 to 5.5. If we add the effect
of such an $R_V$ change on the star count $A_V$ estimates (which needs to be
multiplied by a factor 1.12 due to the fact that our star counts are based on
a $B$ image, see section 2.1), the extinction discrepancy between SFD98 and
star counts, a factor of about 2, is reduced by only a factor 1.34. This paper
thus suggests that the $R_V$-dependent extinction curves should separate again
in the FIR.

\subsection{Validity of the SFD98 extinction map} 
 
\begin{figure}[htb] 
	\includegraphics[width=8.8cm]{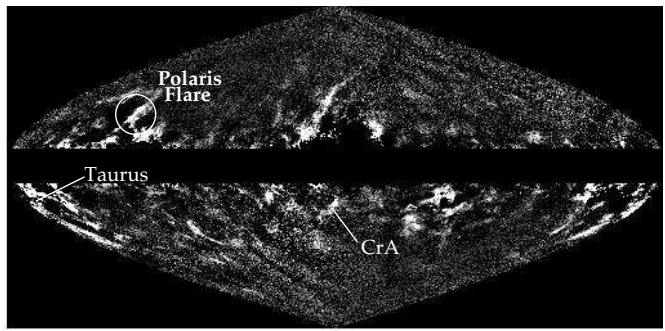} 
\caption{{\em Cold} component of the dust emission at 240~$\mu$m from
\citet{LABP98}} 
\label{cold_emission} 
\end{figure}

We chose the Polaris Flare for its brightness in the {\em cold} emission map 
presented in Fig.~\ref{cold_emission}. The white regions in this figure are
the regions with {\em cold} emission, outside the Galactic plane, where the
present work questions the validity of the SFD98.  
 
As mentioned in Sect.~\ref{add_evid}, additional studies similar to the
present work support our results \citep{SG99,AG99a,vM01} for pieces of the
Corona Australis and Taurus clouds. Both regions have an important
{\em cold} emission in Fig.~\ref{cold_emission}. 
Since the \citet{SG99} work is a prelude to what can be done systematically
with the {\em Sloan Digitized Sky Survey} (SDSS), it is crucial for further
works to be aware that the dust emissivity varies in the FIR, even
for cirrus cloud. \citet{AG99a} suggest that the discrepancy results from a
calibration bias. In the present paper we show that there is no general
calibration that could be used in all directions. Optical properties of grains
vary and the conversion from FIR fluxes to opacity cannot be limited
to a single scaling factor. 
 
For low galactic latitude ($|b| < 10\degr$) it is obvious that the SFD98 map 
has to be used with caution since very different physical environments are 
mixed along the line of sight. 
\citet{CFT+99} propose a 3-dimensional extinction model to constrain the
galactic structure based on the SFD98 map used toward globular clusters as
close as 3\degr\ to the Galactic center. They found a recalibration was
better to fit their data and they used the new calibration for the whole sky.
The Galactic plane and even more, the Galactic center, are very complicated
regions: the presence along the line of sight of massive stars which heat
the interstellar dust and molecular clouds with {\em cold} dust makes
impossible the separation of the different components which contribute
to the FIR flux. It is not clear how their empirical calibration
of the FIR emissivity can be related to those determined for the
nearby interstellar medium.  
 
\citet{BDA99} have derived the FIR dust emissivity using wavelength
dependences derived from FIR spectra of galactic emission and
the SFD98 extinction map to normalize the $\tau_{\rm FIR}/A_V$ ratio.
The authors admit that two components of temperature could exist at low
galactic latitude but they omit that dust properties could also vary. 
 
The knowledge of the extinction close to the Galactic plane is required to
estimate precisely the properties of some objects, like the RR Lyrae type
stars, used as a distance indicator. \citet{SPG99} used the SFD98 map to
deredden RR Lyrae type stars in Baade's window, i.e. 3\degr\ from the
Galactic center.

\section{Conclusion} 
\label{conclusion} 
 
We have presented an extinction map of the Polaris high latitude cloud
derived from star counts. This extinction map is compared with that produced
by SFD98 based on the IRAS and DIRBE FIR sky maps. Within the Polaris
cirrus cloud, the SFD98 extinction value is found to be a factor 2 to 3
higher than the star count values. 
 
We propose to relate the extinction discrepancy to variations in the
$\tau_{\rm FIR}/A_V$ of interstellar dust from the {\em warm } and {\em cold}
emission components introduced by \citet{LABP98} on the basis of the observed
difference in small grain abundance and large grain temperature between the
diffuse ISM and dense molecular gas. Within this interpretation, we find that
the FIR emissivity of dust grains in the {\em cold} component is on average
about 4.0 times (from 2.7 to 6.0) larger than it is in the {\em warm}
component. 
There is a slight evidence that the evolution of the grain properties are
gradual, in the sense that the $\tau_{\rm FIR}/A_V$ ratio increases
when the {\em cold} dust temperature decreases. The increase in the FIR
opacity could be tracing a growth of dust grains by coagulation. Our work
is thus suggesting that the size/porosity of the large dust grains evolve
within low opacity molecular cloud, such as the Polaris Flare. 

SFD98 have converted the FIR opacity in extinction assuming a single emission
temperature for each line of sight and $\tau_{\rm FIR}/A_V$ ratio. 
Our work questions the validity of this assumption in all regions
with significant {\em cold} emission (Fig.~\ref{cold_emission}). A few
independent studies in other molecular clouds support this statement.
The SFD98 extinction could be overestimating the true extinction in regions
with {\em cold} emission that are found even at high Galactic latitude and 
for extinction as low as $A_V \gtrsim 0.5$.
Variations in the dust emissivity preclude the conversion from FIR optical
depth to $A_V$ with one single factor valid over the whole sky. 
Low latitude regions ($b<10^\circ$) and a small fraction of the high latitude
sky require special attention.

\begin{acknowledgements} 
L. Cambr\'esy acknowledges partial support from the Lavoisier grant 
of the French Ministry of Foreign Affairs. 
\end{acknowledgements}

\end{document}